# An anomalous 'butterfly'-shaped magnetoresistance loop in an alloy, $Tb_4LuSi_3$

K. Mukherjee, Sitikantha D Das, Niharika Mohapatra, Kartik K Iyer and E.V. Sampathkumaran[*]
*Tata Institute of Fundamental Research, Homi Bhabha Road, Colaba, Mumbai 400005, India*

Magnetic field ($H$) induced first-order magnetic transition and the associated electronic phase-separation phenomena are active topics of research in magnetism. Magnetoresistance (MR) is a key property to probe these phenomena and, in literature, a butterfly-shaped MR loop has been noted while cycling the field, with the envelope curve lying *below* the virgin curve in MR versus $H$ plots of such materials. Here, we report an opposite behavior of MR loop for an alloy, $Tb_4LuSi_3$, at low temperatures (<<20 K) in the magnetically ordered state. Such an anomalous curve reveals *unexpected domination of higher resistive* high-field phase in electrical conduction, unlike in other materials where conduction is naturally by *low-resistive* high-field phase that follows first-order transition. The observed features reveal an unusual electronic phase separation, namely involving *high-resistive high-field phase* and low-resistive virgin phase.

The phenomenon of magnetic phase co-existence following a travel through a first-order metamagnetic transition has been actively studied for more than a decade in many materials, particularly in the context of the physics of manganites [1]. Generally speaking, in all such magnetic materials known to date, at a given temperature ($T$), an externally applied magnetic field ($H$) transforms 'less electrically conductive' antiferromagnetic phase to a 'more conductive' ferromagnetic phase at the first-order transition, resulting in negative magnetoresistance (MR defined as $\{\rho(H)-\rho(0)\}/\rho(0)$, where $\rho$ is electrical resistivity). If the magnetic field is gradually reduced to zero, the variation of $\rho$ with $H$ can be hysteretic, in which case a lower value of $\rho$ compared to that for the virgin state has been naturally observed with the value of $\rho$ after returning the field to zero depending on the fractions of this 'supercooled' high-field phase and 'transformed' virgin phase contributing to electrical conductivity. The above-stated variation of $\rho$ with $H$ is found to be true irrespective of the nature of the magnetic interaction mediating magnetic ordering, that is, whether it is double-exchange mechanism as in manganites or Rudermann Kasuya Kittel Yosida interaction as in rare-earth intermetallics [see, for instance, Refs. 3, 4].

Recently, we have reported [5, 6] that the compound, $Tb_5Si_3$, crystallizing in $Mn_5Si_3$-type hexagonal structure (space group $P6_3/mcm$) [7-9] interestingly attains a higher-resistive state beyond a critical magnetic field ($H_{cr}$) in the magnetically ordered state (<70 K), in contrast to commonly known behavior in metamagnetic systems. In this article, we report that, for a partial replacement of Tb by Lu (20% atomic percent), the electrical transport in zero-field, attained after traveling through $H_{cr}$ once, is dominated by the "supercooled" high-resistive state interestingly *resulting in the virgin ρ(H) curve falling below the envelope curve in the entire field range of investigation even in the negative cycles of H*. The results reveal that this system provides a unique opportunity to study super-cooling and electronic phase separation phenomena for a case in which high-field phase is *less conductive*. We have also studied a few other compositions in the series, $Tb_{5-x}Lu_xSi_3$ ($x=$ 2 and 3), to bring out that this transport behavior is unique to this alloy.

Polycrystalline samples, $Tb_{5-x}Lu_xSi_3$ ($x=$ 1, 2 and 3), were prepared by arc melting stoichiometric amounts of high purity (>99.9%) constituent elements in an atmosphere of high purity argon. Single phase nature and homogeneity of the specimens were ascertained by x-ray diffraction (Cu K$_\alpha$) (figure 1), scanning electron microscope and energy dispersive x-ray analysis. A comparison of x-ray diffraction patterns of the parent and Lu substituted alloys is made in figure 1; this reveals a gradual shift of diffraction lines with Lu substitution thereby establishing that all Lu indeed go to Tb site without precipitating any other phase within the detectable limits of this technique. The $\rho$ measurements in the presence of magnetic fields (<120 kOe, $T=$ 1.8-300 K) were performed by a commercial physical property measurements system (PPMS) (Quantum Design) and a conducting silver paint was used for making electrical contacts of the leads with the samples. We had to characterize the specimens by dc magnetization, $M$, (<120 kOe, $T=$ 1.8-300 K) for a comparison with the transport behavior and this was done with the help of a commercial vibrating sample magnetometer (Oxford Instruments).

We first look at how the magnetic anomalies vary with a gradual replacement of Tb by Lu. In figure 2, we show magnetization measured in a field of 5 kOe as a function of temperature for all compositions. The data for the parent compound from our past publications [5, 6] is included for the benefit of the reader. As expected, the magnetic transition, as indicated by the peak temperature in *M/H* plots obtained in a field of 5 kOe, shifts to lower temperatures monotonically with increasing Lu concentration. The $M(H)$ plots (see figure 3) undergo dramatic changes in the magnetically ordered state with Lu substitution. For instance, for $x= 1.0$, at 1.8 K, the field-induced transition is feeble and significantly broadened, and a continuous increase in slope (rather than an abrupt one reported for $x= 0$ near 58 kOe, see inset of figure 3) in $M(H)$ plot *beyond 20 kOe* is noted in the increasing field direction. This feature is absent in the reverse leg of the $M(H)$ curves. [The virgin curve lies outside the envelope curve as shown later, thereby suggesting that the field-induced transition is of a first-order character type, but broadened]. The change of slope was found to get further weakened as the temperature increases (not shown here). For higher concentrations of Lu (figure 3), the variation of $M$ with $H$ in the magnetically ordered state (e.g., at 1.8 K) does not reveal any spin reorientation effects. It is important to note that, for the $x = 1.0$ alloy, the value of $M$ at the highest field measured (about 17.5 $\mu_B$/formula-unit at 120 kOe) is nearly the same as that obtained by linear extrapolation of the low-field data (that is, before the transition, < 40 kOe) of $Tb_5Si_3$. This could mean that only some portion of Tb ions, possibly decided by chemical inhomogeneity resulting from Lu substitution, undergo the magnetic transition at $H_{cr}$ in the alloy, $Tb_4LuSi_3$, and this explains why the transition is feeble.

Let us look at the MR behavior (figure 4). For $x= 1$ at 1.8 K, we see a fairly prominent upturn in the range 30-50 kOe in the virgin curve followed by a decrease at higher fields as in the parent compound [5, 6] (see inset of figure 4). This transition is observed despite the fact that it is weak and broadened in $M/H$. This means that the number of Tb ions undergoing this transition is sufficient enough to provide percolative electrical conduction. The fact that the fraction of Tb ions undergoing field-induced magnetism is diminished compared to that in the parent compound could be qualitatively inferred from a relatively reduced jump (about 60%) in MR at $H_{cr}$. The transition field is reduced with respect to that in the parent compound (from ~ 58 to ~ 50 kOe). Apart from dilution effect of Tb sublattice, we believe that positive pressure also is responsible for this reduction based on our experiments under external pressure and negative chemical pressure induced by Ge substitution for Si [6, 10]. For higher concentrations of Lu, the features due to field-induced transition are not apparent (figure 4), possibly because $H_{cr}$ is reduced to zero due to these factors. In fact, MR remains in the negative zone in the entire field range of investigation without any evidence for hysteresis. We will make more comments on the magnetic behavior of these alloys later in this article. The point of emphasis here is that, among the compositions we studied, the alloy, $Tb_4LuSi_3$, is the one of importance for the present purpose.

Let us now look at the MR behavior while returning the field to zero after reaching 120 kOe for $x= 1$ to infer the nature of the high-field (and supercooled) phase. MR keeps increasing closely following the virgin curve till about 60 kOe and, at lower fields, the curve diverges from the virgin curve with this increasing trend persisting till the field is reduced to zero (as though there is an extrapolation of the high-field phase behavior). This situation is different from the parent compound in the sense that the increase in $\rho$ in this case is cut off by a sharp fall before the field reaches zero (see inset of figure 4). The value of MR in zero field thus attained for the former is relatively larger (about 50%). If one has to observe the increasing tendency till zero field for the parent compound, an external pressure needs to be applied at 1.8 K [6]. It is important to note that MR at low fields increases essentially quadratically (see a broken line in figure 4) with decreasing *H characteristic of paramagnets.* We have earlier mooted [6, 10] the idea of 'inverse metamagnetism (a process in which paramagnetic fluctuations are induced at $H_{cr}$) to explain sudden enhancement of positive MR in the parent compound at $H_{cr}$. Such an 'inverse' process can happen in a situation in which the molecular field due to one magnetic site induces an antiferromagnetic component at the other site and an application of an external magnetic field (at a critical value) tends to destroy this coupling thereby resulting in magnetic fluctuations (and hence increased scattering). Clearly, if such 'a high-field phase' with magnetic fluctuations is 'supercooled' to zero-field, one should see quadratic field-dependence of MR as $H \rightarrow 0$, as observed experimentally.

In view of the exotic MR behavior of $Tb_4LuSi_3$ stabilized under ambient pressure conditions as described above, we considered it worthwhile to perform additional isothermal MR experiments for this composition traveling through negative values of $H$ to emphasize on the key conclusion. We have noted that there is some degree of hysteresis of isothermal $M$ curve persisting even at 120 kOe (see figure 5a for 1.8 K data ), but the size of the loop was found to get weaker gradually with increasing temperature. The location of the virgin curve outside envelope curve is distinctly visible as a typical feature of broadened field-induced first-order magnetic transitions. In figure 5b, we show MR data for both positive and negative cycles of $H$ at 1.8, 10 and 25 K. Arrows and numericals are placed on the curves to serve as guides to the eyes. It is apparent from this figure that, at 1.8 K, while increasing the magnitude of the field in the negative $H$ quadrant, there is a monotonic decrease of MR *without any evidence for the field-induced transition* (as though the conductivity occurs through the supercooled phase only). With the consideration of the data for further cycling of magnetic-field, a butterfly-shaped MR curve is evident with the virgin curve lying *below* this envelope curve. *The observation of this shape of MR loop is unique in the field of magnetism.* With increasing temperature, say to 10 K, in the positive quadrant, MR in the reverse leg tends to fall at a particular field (< ~25 kOe) at which the supercooled state tends to get transformed to the virgin state. In the zero-field reached thereafter, MR stays 'intermediate' between that expected for the virgin phase and the high-field phase. This implies that, after traveling through the transition field, at this temperature, the fraction of the high-field phase in zero-field gets reduced with respect to that for 1.8 K. As a further support for the gradual dominance of virgin phase following field-cycling at 10 K, there is an upturn in MR near -40 kOe (as in virgin curve), however, with a *reduced magnitude* compared to that for virgin state. A similar *reduced* jump appears again in the positive quadrant for further field cycling. Clearly, the virgin curve lies below this butterfly-shaped MR curve. The data at 10 K distinctly brings out that there is an unusual phase-coexistence involving high-field-high-resistive phase and low-field-low-resistive phase after the returning the field to zero. At 25 K, there is a field-induced transition near 25 kOe and the ρ value in zero field after traveling through this field is nearly the same as that of the virgin curve. Thus, one is able to control the fraction of the virgin phase and the high-field phase by varying the temperature. Unfortunately, one can not obtain the relative fractions of these two phases from the corresponding isothermal magnetization curves in the event that the supercooled component is of a paramagnetic-like fluctuating phase as argued earlier [6, 10].

Now to bring out the uniqueness of the MR behavior of $x$= 1 alloy among these compositions, we make relevant comments on the magnetic behavior of other Lu rich compositions, $x$= 2 and 3. As mentioned earlier, as Lu concentration increases, $H_{cr}$, is presumably reduced zero. This means that, in these Lu richer alloys, these Tb ions should show the MR behavior of the high-field phase of $x$= 0 or 1 alloys, that is, a gradual drop in ρ with an increase $H$. This is indeed found to be the case (see figure 4). In fact, the MR curves for these higher compositions of Tb (at 1.8 K) look somewhat similar to that in the reverse leg of MR for $x$= 1 alloy in the sense that MR varies with $H$ essentially quadratically (as shown in the bottom part of figure 4). Such an $H$-dependence of MR is a characteristic of paramagnetic-like fluctuations alone and not of magnetically ordered state. The MR curves were found to be symmetric with respect to zero field without any hysteresis. However, the *M*(*H*) curves are hysteretic at 1.8 K as shown in figure 3. The hysteretic *M*(*H*) with a gradual variation with *H* without any evidence for saturation at high-fields implies a complex antiferromagnetic component. Apart from this *M*(*H*) behavior, in figure 2, we note clear evidence for magnetic ordering in *M(T)* data. Thus, there appears to be a conflict in the conclusions from MR on the one hand and *M* on the other. The only way to reconcile these apparently conflicting inferences in these single-phase materials is to propose that, even for these compositions, there is an electronic phase separation due to chemical inhomogeneity. This means that, to start with (that is, in the virgin state of these compositions), there is a paramagnetic-like region responsible for MR behavior, coexisting with the magnetic region which does not dominate conductivity. Thus, this family of alloys is in general ideal to study the novel electronic phase separation in a metallic environment. Incidentally, the high-field magnetic phase undergoes changes with increasing *x* is evident from the fact that the magnetization value (per Tb) at 120 kOe varies non-monotonically with decreasing Tb concentration.

    Summarizing, the magnetoresistance behavior of $Tb_4LuSi_3$ is exceptional in magnetism. That is, the magnetoresistance versus magnetic field loop for this compound exhibits butterfly-shaped behavior *with the virgin curve lying lower with respect to envelope curve.* We have demonstrated that such a shape of MR curve can arise in the event that the high-field phase following field-induced first-order magnetic transition is (unexpectedly) more resistive electrically compared to virgin magnetic phase and that it dominates conductivity in subsequent field-cycling. The present study brings out an opportunity to probe an unusual electronic phase separation.

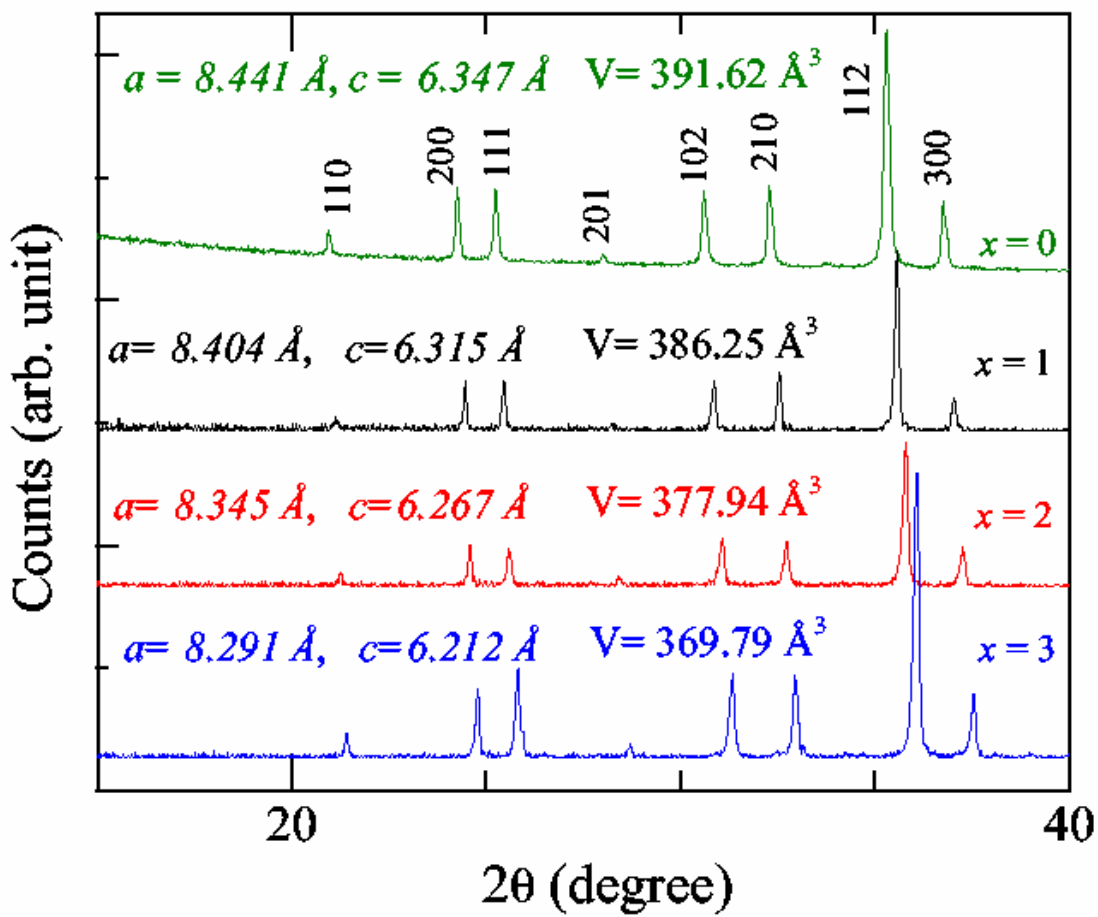

Figure 1:
(color online) X-ray diffraction patterns below $2\theta = 40°$ for the alloys, $Tb_{5-x}Lu_xSi_3$. The lattice constants, $a$ and $c$ ($\pm 0.004$ Å) and unit-cell volume (V) are included. The curves are shifted along $y$-axis for the sake of clarity.

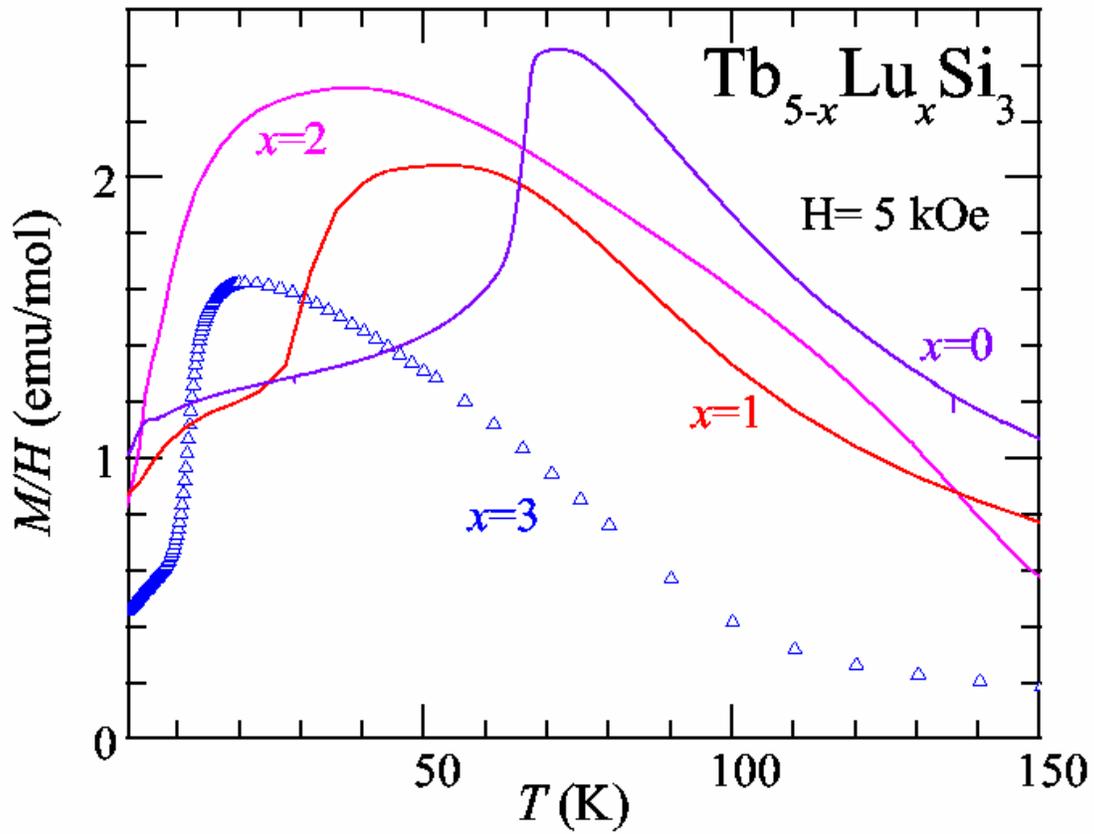

Figure 2:
(color online) Magnetization divided by magnetic field as a function of temperature obtained in a field of 5 kOe for the alloys, $Tb_{5-x}Lu_xSi_3$ ($x$= 0, 1, 2, and 3). The data points are shown for $x$= 3 only.

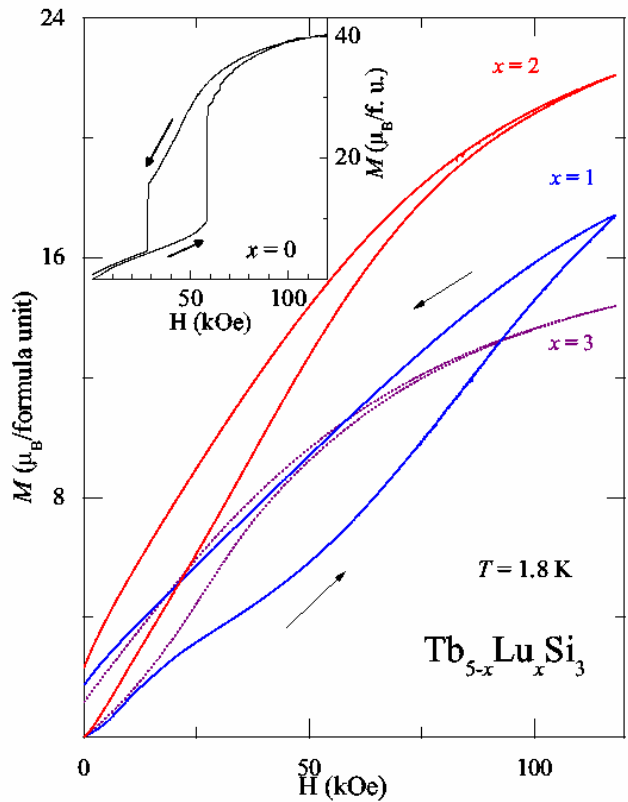

Figure 3:
(color online) Isothermal magnetization at 1.8 K for the alloys, $Tb_{5-x}Lu_xSi_3$. The curve for $Tb_5Si_3$ (Ref. 5) is shown in the inset.

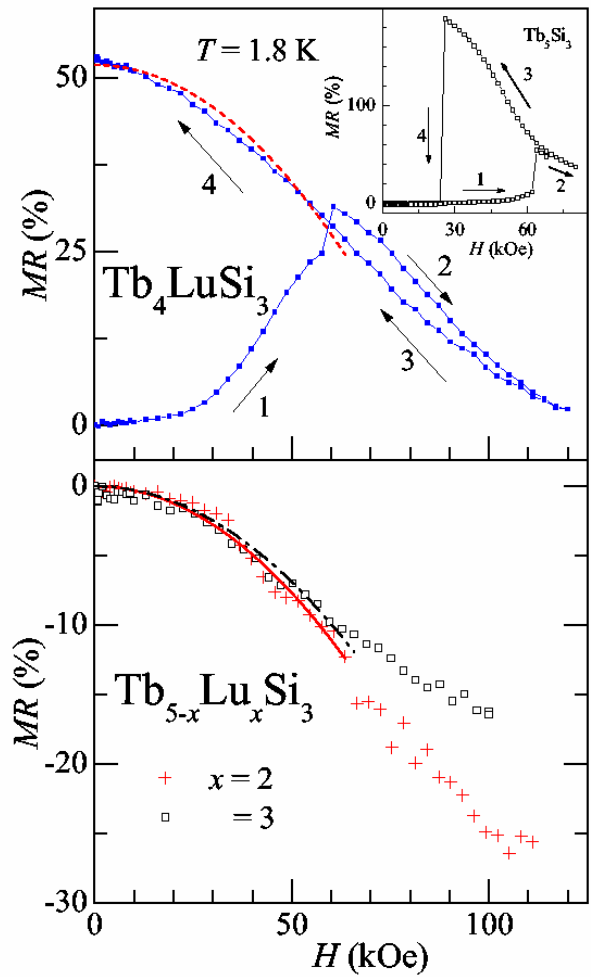

Figure 4:
(color online) Magnetoresistance as a function of externally applied magnetic field for the alloys, $Tb_{5-x}Lu_xSi_3$ at 1.8K. Lines are drawn through the data points for $Tb_4LuSi_3$. A dotted line is drawn in the reverse field-cycle for this composition to highlight that MR varies essentially quadratically with $H$. Continuous lines for other compositions represent quadratic field dependence. Arrows and numericals (top figure) are drawn as a guide to the eyes.

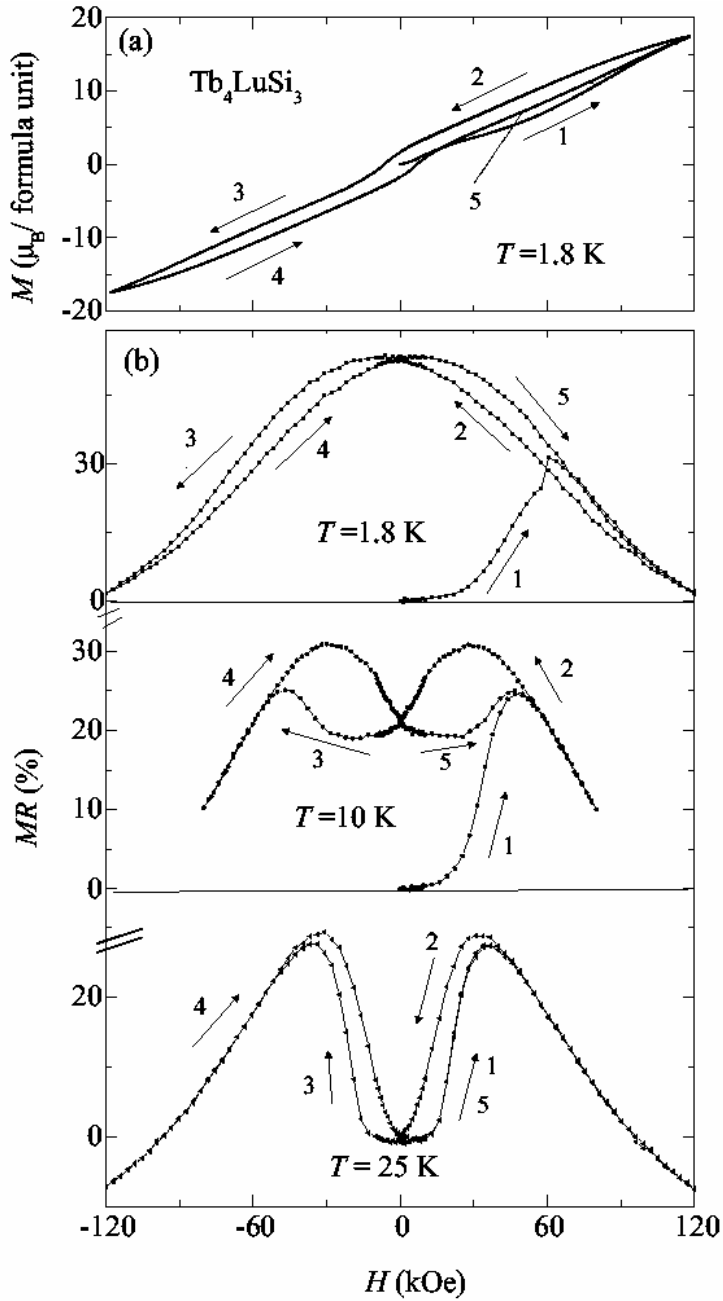

Figure 5:
(a) Isothermal magnetization at 1.8 K and (b) magnetoresistance at 1.8, 10 and 25 K for $Tb_4LuSi_3$. The lines through the data points and arrows and numericals are drawn as a guide to the eyes.